\documentclass[10pt, twocolumn, pre, aps, superscriptaddress, showpacs]{revtex4-1}
\usepackage{amsmath, graphicx, subfigure, tikz}
\begin{document}

    \title{Phase Transitions between Different Spin-Glass Phases and\\ between Different Chaoses in Quenched Random Chiral Systems}
    \author{Tolga \c{C}a\u{g}lar}
    \affiliation{Faculty of Engineering and Natural Sciences, Sabanc\i\ University, Tuzla, Istanbul 34956, Turkey}
    \author{A. Nihat Berker}
    \affiliation{Faculty of Engineering and Natural Sciences, Kadir Has University, Cibali, Istanbul 34083, Turkey}
    \affiliation{Department of Physics, Massachusetts Institute of Technology, Cambridge, Massachusetts 02139, USA}

%05.10.Cc    Renormalization group methods
%64.60.ae    Renormalization-group theory
%64.60.De    Statistical mechanics of model systems
%75.10.Nr    Spin-glass and other random models
%75.50.Lk    Spin glasses and other random magnets

%64.60.Cn    Order-disorder transformations
%05.50.+q    Lattice theory and statistics (Ising, Potts, etc.)
%61.43.-j    Disordered solids
%75.10.Hk    Classical spin models

\begin{abstract}

The left-right chiral and ferromagnetic-antiferromagnetic double
spin-glass clock model, with the crucially even number of states
$q=4$ and in three dimensions $d=3$, has been studied by
renormalization-group theory. We find, for the first time to our
knowledge, four different spin-glass phases, including conventional,
chiral, and quadrupolar spin-glass phases, and phase transitions
between spin-glass phases. The chaoses, in the different spin-glass
phases and in the phase transitions of the spin-glass phases with
the other spin-glass phases, with the non-spin-glass ordered phases,
and with the disordered phase, are determined and quantified by
Lyapunov exponents.  It is seen that the chiral spin-glass phase is
the most chaotic spin-glass phase. The calculated phase diagram is
also otherwise very rich, including regular and temperature-inverted
devil's staircases and reentrances.

\end{abstract}
\maketitle

\section{Introduction}

Spin-glass phases, created by competing frustrated random
ferromagnetic and antiferromagnetic interactions, have been known
\cite{NishimoriBook} to incorporate a plethora of interesting
complex phenomena, not the least being the natural generation chaos
\cite{McKayChaos, McKayChaos2,BerkerMcKay}. Recently, it has been
shown \cite{Caglar, Caglar2} that competing left- and right-chiral
interactions also create spin-glass phases, even in the absence of
competing ferromagnetic and antiferromagnetic interactions.  First
shown \cite{Caglar} with chiral Potts models \cite{Ostlund, Kardar,
Huse, Huse2, Caflisch} with the inclusion of quenched randomness,
chiral spin glasses were recently extended \cite{Caglar2} to clock
models with an odd number of states $(q=5)$, resulting in a
literally moviesque sequence of phase diagrams, including regular
and inverted devil's staircases, a chiral spin-glass phase, and
algebraic order.

The chiral clock model work was purposefully initiated
\cite{Caglar2} with odd number of states $q$, in order to deal with
the complexity of the global phase diagram, since it is known that
the odd $q$ models do not show \cite{Ilker3} the traditional
ferromagnetic-antiferromagnetic spin-glass phase. This is because
neighboring, antiferromagnetically interacting odd $q$ clock spins
cannot achieve perfect antiferromagnetic alignment. Furthermore,
there are two configurations for the near-antiferromagnetic
alignment, creating a built-in disorder. The traditional
ferromagnetic-antiferromagnetic spin-glass phase does not occur and
the antiferromagnetic phase is a critical phase lacking conventional
long-range order.\cite{Ilker3} On the other hand, the even $q$ clock
spins can achieve complete antiferromagnetic pairing, and exhibit
the conventional antiferromagnetic long-range order and the
traditional ferromagnetic-antiferromagnetic spin-glass
phase.\cite{Ilker1} Thus, the current study is on the random chiral
clock model with an even number of states $(q=4)$, which supports
the ferromagnetic-antiferromagnetic usual spin-glass phase
\cite{Ilker1}, as well as, as we shall see below, with added phase
diagram complexity, the chiral spin-glass phase and two other new
spin-glass phases. A double spin-glass model is constructed,
including competing quenched random left-right chiral and
ferromagnetic-antiferromagnetic interactions, and solved in three
dimensions by renormalization-group theory.

The extremely rich phase diagram includes, to our knowledge for the
first time, more than one (four) spin-glass phases on the same phase
diagram and three separate spin-glass-to-spin-glass phase
transitions. These constitute phase transitions between chaoses.  We
determine the chaotic behaviors of the spin-glass phases, of the
phase transitions between the spin-glass phases, of the phase
transitions between the spin-glass phases and the ferromagnetic,
antiferromagnetic, quadrupolar, and disordered phases.

\section{Doubly Spin Glass System:\\ Left-Right Chiral and Ferro-Antiferro Interactions}
The \textbf{$q-$state clock spin glass} is composed of unit spins
that are confined to a plane and that can only point along $q$
angularly equidistant directions, with Hamiltonian
\begin{equation}
-\beta {\cal H} = \sum_{\left<ij\right>} J_{ij} \vec{s}_i.\vec{s}_j
= \sum_{\left<ij\right>} J_{ij}\cos\theta_{ij}, \label{eq:qclockBH}
\end{equation}
where $\beta = 1/k_BT$, $\theta_{ij} = \theta_i - \theta_j$, at each
site $i$ the spin angle $\theta_i$ takes on the values
$(2\pi/q)\sigma_i$ with $\sigma_i=0,1,2,\ldots,(q-1)$, and
$\left<ij\right>$ denotes summation over all nearest-neighbor pairs
of sites. As the long-studied ferromagnetic-antiferromagnetic
spin-glass system \cite{NishimoriBook}, the bond strengths $J_{ij}$,
with quenched (frozen) ferromagnetic-antiferromagnetic randomness,
are $+J > 0$ (ferromagnetic) with probability $1-p$ and $-J$
(antiferromagnetic) with probability $p$, with $0 \leq p \leq 1$.
Thus, the ferromagnetic and antiferromagnetic interactions locally
compete in frustration centers.  Recent studies on
ferromagnetic-antiferromagnetic clock spin glasses are in Refs.
\cite{Ilker1,Ilker3,Lupo}.

In the \textbf{$q-$state chiral clock double spin glass}, recently
introduced (and used in the qualitatively different odd $q = 5$),
frustration also occurs via randomly frozen left or right chirality
\cite{Caglar}, thus doubling the spin-glass mechanisms. The
Hamiltonian in Eq. (1) is generalized to random local chirality,
\begin{equation}
\label{eq:qccsgBH} -\beta {\cal H} = \sum_{\left<ij\right>} [
J_{ij}\cos\theta_{ij} + \Delta \, \delta(\theta_{ij} + \eta_{ij}
\frac{2\pi}{q})].
\end{equation}
In a cubic lattice, as sites along the respective coordinate
directions are considered, the $x,y,$ or $z$ coordinates increase.
Since bond moving in the Migdal-Kadanoff approximation
\cite{Migdal,Kadanoff} is done transversely to the bond directions,
this sequencing is respected. Equivalently, in the corresponding
hierarchical lattice
\cite{BerkerOstlund,Kaufman1,Kaufman2,McKay,Hinczewski1}, one can
always define a direction along the connectivity, for example from
left to right, and assign consecutive increasing number labels to
the sites.  In Eq. (2), for each pair of nearest-neighbor sites
$\left<ij\right>$ the numerical site label $j$ is ahead of $i$,
frozen (quenched) $\eta_{ij} = 1$ (left chirality) or $-1$ (right
chirality), and the delta function $\delta(x)=1\,(0)$ for $x=0\,
(x\neq 0)$. The overall concentrations of left and right chirality
are respectively $1-c$ and $c$, with $0 \leq c \leq 1$. The strength
of the random chiral interaction is $\Delta/J$, with temperature
divided out. Thus, the system is chiral for $\Delta \neq 0$,
chiral-symmetric for $c=0.5$, chiral-symmetry-broken for $c\neq0.5$.
The global phase diagram is in terms of temperature $J^{-1}$,
antiferromagnetic bond concentration $p$, random chirality strength
$\Delta / J$, and chiral symmetry-breaking concentration $c$.

\section{Renormalization-Group Method: Migdal-Kadanoff Approximation and Exact Hierarchical Lattice Solution}

Our method, previously described in extensive detail \cite{Caglar2}
and used on a qualitatively different model with qualitatively
different results, is simultaneously the Migdal-Kadanoff
approximation \cite{Migdal,Kadanoff} for the cubic lattice and the
exact solution
\cite{BerkerOstlund,Kaufman1,Kaufman2,McKay,Hinczewski1} for a $d=3$
hierarchical lattice, with length rescaling factor $b=3$. Exact
calculations on hierarchical lattices are also currently widely used
on a variety of statistical mechanics
\cite{Derrida,Thorpe,Efrat,Monthus2,
Lyra,Xu2014,Silva,Boettcher1,Boettcher2,Hirose2,Boettcher3,Nandy,Boettcher4,
Bleher, Zhang2017}, finance \cite{Sirca}, and, most recently,
DNA-binding \cite{Maji} problems.

\begin{figure*}[ht!]
\centering
\includegraphics[scale=1]{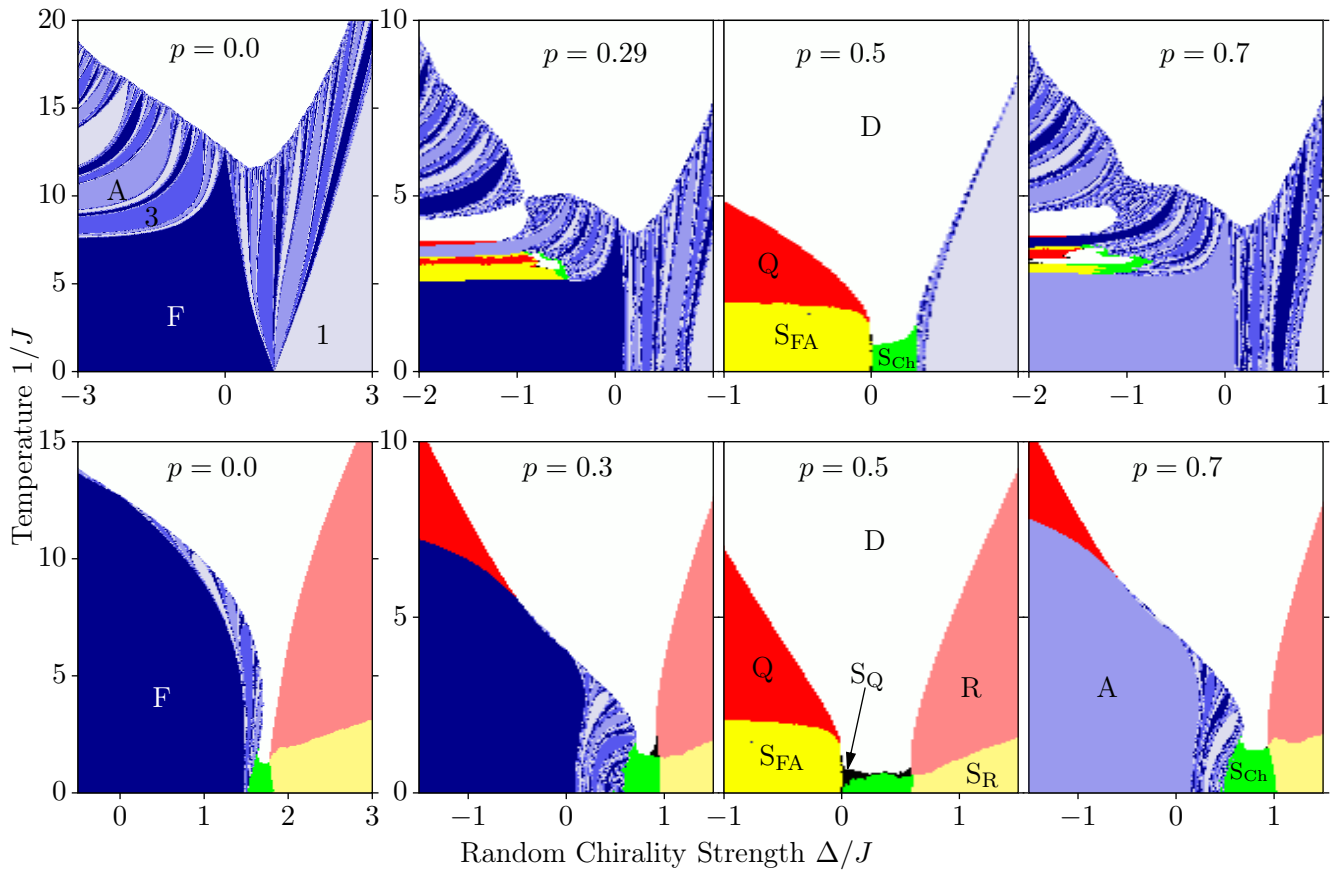}
\caption{(Color online) A calculated sequence of phase diagrams for
the left chiral ($c=0$), on the upper side, and quenched random left
and right chiral ($c=0.5$), on the lower side, systems with, in both
cases, quenched random ferromagnetic and antiferromagnetic
interactions. The horizontal axis is the random chirality strength
$\Delta/J$. The consecutive phase diagrams are for different
concentrations $p$ of antiferromagnetic interactions. The system
exhibits a ferromagnetic phase F, an antiferromagnetic phase A, a
multitude of different chiral phases, a quadrupolar phases Q, a
"one-step" phase R, and four differently ordered spin-glass phases:
the chiral spin-glass S$_{CH}$, the usual
ferromagnetic-antiferromagnetic spin glass S$_{FA}$, the quadrupolar
spin glass S$_Q$, and S$_R$. The phase diagrams obtained from $p$
and $1-p$ are symmetric, since the system has an even number of spin
directions. On some of the chiral phases, the $\pi/2$ multiplicity
of the asymptotically dominant interaction is indicated. The
ferromagnetic and chiral phases accumulate as different devil's
staircases at their boundary with the disordered (D) phase.}
\end{figure*}

\begin{figure}[ht!]
\centering
\includegraphics[scale=1]{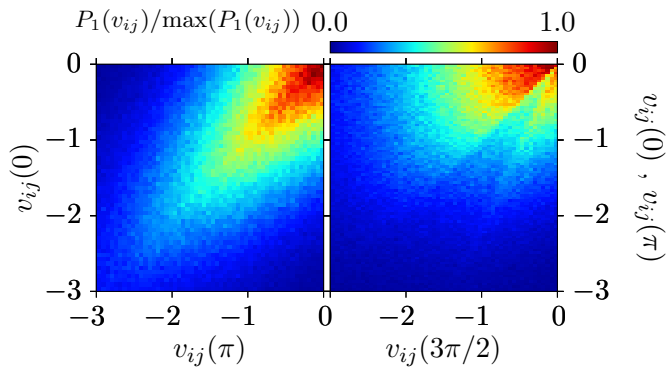}
\caption{\label{fig:chsg} (Color online) Asymptotic fixed
distribution of the chiral spin-glass phase S$_{Ch}$. The part of
the fixed distribution $P_1({\bf V}_{ij})$, for the interactions
${\bf V}_{ij}$ in which $V_{ij}(\pi/2)$ is maximum and therefore $0$
(and the other three interactions are negative) is shown in this
figure, with $v_{ij}(\theta) =
V_{ij}(\theta)/\left<\left|V_{ij}(\theta)\right|\right>$. The
projections of $P_1({\bf V}_{ij})$ onto two of its three arguments
are shown in each panel of this figure. The other three
$P_\sigma({\bf V}_{ij})$ have the same fixed distribution. Thus
chirality is broken locally but not globally, just as, in the
long-time studied ferromagnetic-antiferromagnetic spin glasses,
spin-direction symmetry breaking is local but not global (i.e., the
local magnetization is non-zero, the global magnetization is zero).}
\end{figure}

\begin{figure}[ht!]
\centering
\includegraphics[scale=0.9]{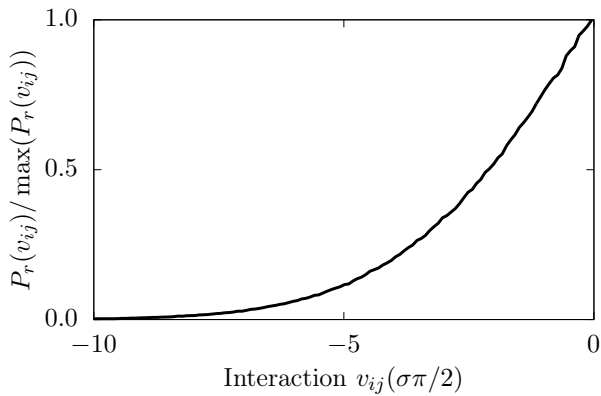}
\caption{\label{fig:nsg} Asymptotic fixed distributions of 3
different spin-glass phases, with $v_{ij}(\theta) =
V_{ij}(\theta)/\left<\left|V_{ij}(\theta)\right|\right>$.  For the
ferromagnetic-antiferromagnetic spin-glass S$_{FA}$ phase, $r=0,
\sigma=2$ and $r=2, \sigma=0$.  The other two angles do not occur.
For the quadrupolar spin-glass S$_Q$ phase, $r=0, \sigma=1$ and
$r=1, \sigma=0$, with $V_{ij}(0) = V_{ij}(\pi)$ and $V_{ij}(\pi/2) =
V_{ij}(3\pi/2)$. For the spin-glass S$_R$ phase, $r=1, \sigma=3$ and
$r=3, \sigma=1$. The other two angles do not occur. The $v_{ij}(0) =
v_{ij}(\pi)$ curve obtained from the left panel of Fig. 2 also
matches the curve here.}
\end{figure}

\begin{figure*}[ht!]
\centering
\includegraphics[scale=1]{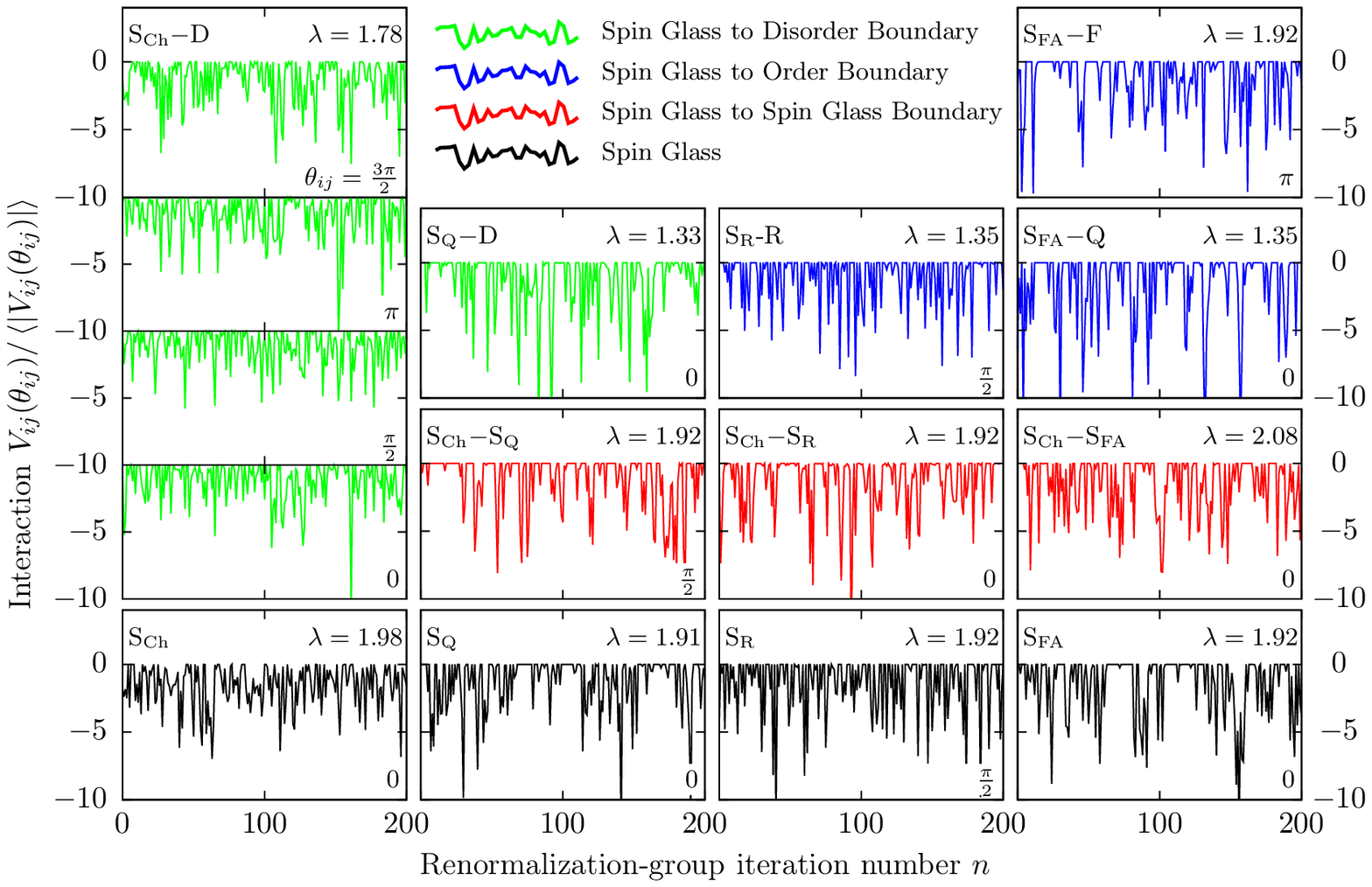}
\caption{\label{fig:traj}Chaotic renormalization-group trajectories
of the four different spin-glass phases (black), of the phase
boundaries of the spin-glass phases with other spin-glass phases
(red), with non-spin-glass ordered (blue) and disordered (green)
phases. The phase boundary chaoses of each spin-glass phase are
given in their corresponding vertically aligned panels. In each
case, only one of the four interactions $V_{ij}(0), V_{ij}(\pi/2),
V_{ij}(\pi), V_{ij}(3\pi/2)$ at a given location $\left<ij\right>$,
under consecutive renormalization-group transformations, is shown,
except, for illustration, all four interactions are shown for the
chaos at the phase transition between the chiral spin-glass and
disordered phases. The $\theta_{ij}$ angular value of each
interaction $V_{ij}(\theta_{ij})$ is indicated in the figure panels,
as well as the Lyapunov exponent $\lambda$ calculated from the
chaotic sequence under renormalization-group transformations. The
Lyapunov exponent is calculated over 1,000 renormalization-group
iterations, after throwing out the first 200 iterations. Inside all
four spin-glass phases, the average interaction diverges as $<|V|>
\sim b^{y_R n}$, where $n$ is the number of renormalization-group
iterations and $y_R = 0.25$ is the runaway exponent.  At the
S$_{Ch}-$S$_R$, S$_{Ch}-$S$_Q$, S$_{FA}-$A, S$_{FA}-$F phase
boundaries, $y_R = 0.25$ also.  At the S$_{Ch}-$S$_{FA}$ phase
boundary, $y_R = 0.11$ for $V_{ij}(0), V_{ij}(\pi)$ and $y_R = 0.25$
for $V_{ij}(\pi/2), V_{ij}(3\pi/2)$. At the phase boundaries of the
spin-glass phases with some non-spin-glass-ordered and disordered
phases, the average interaction remains non-divergent, fixed at $<V>
= -0.34$ for S$_{FA}-$Q, S$_R-$R, S$_Q-$D and $<V> = -1.07$ for
S$_{Ch}-$D. As indicated by the Lyapunov exponents, chaos is
stronger inside the chiral spin-glass phase.}
\end{figure*}

Under the renormalization-group transformation \cite{Caglar2}, the
Hamiltonian of Eq. (2) maps onto the more general form
\begin{equation}
-\beta {\cal H} = \sum_{\left<ij\right>} V_{ij}(\theta_{ij}),
\label{eq:renhamil}
\end{equation}
where $\theta_{ij} = \theta_i - \theta_j$ can take $q = 4$ different
values, so that for each pair $<ij>$ of nearest-neighbor sites,
there are q = 4 different interaction constants
\begin{multline}
\{V_{ij}(\theta_{ij})\} = \\
\{V_{ij}(0),V_{ij}(\pi/2),V_{ij}(\pi),V_{ij}(3\pi/2)\} \equiv
\textbf{V}_{ij} ,
\end{multline}
which are in general different at each locality (quenched
randomness). The largest element of $\{V_{ij}(\theta_{ij})\}$ at
each locality $<ij>$ is set to zero, by subtracting the same
constant from all $q = 4$ interaction constants, with no effect on
the physics; thus, the $q-1 = 3$ other interaction constants are
negative.

The starting double-bimodal quenched probability distribution of the
interactions, characterized by $p$ and $c$ as described above, is
not conserved under rescaling. The renormalized quenched probability
distribution of the interactions is obtained by the convolution
\cite{Andelman}
\begin{equation}
P'(\textbf{V}'_{i'j'}) =
\int{\left\{\prod_{ij}^{i'j'}d\textbf{V}_{ij}P(\textbf{V}_{ij})\right\}\delta(\textbf{V}'_{i'j'}
- \textbf{R}(\left\{\textbf{V}_{ij}\right\}))}, \label{eq:conv}
\end{equation}
where $\textbf{V}_{ij} \equiv \{V_{ij}(\theta_{ij})\}$ as in Eq.
(4), $\textbf{R}(\{\textbf{V}_{ij}\})$ represents the
renormalization-group recursion relation \cite{Caglar2}, primes
refer to the renormalized system, and the procedure is effected
numerically. The different phases and phase transitions of the
system are identified by the different asymptotic
renormalization-group flows of the quenched probability distribution
$P(\textbf{V}_{ij})$.  Similar previous studies, on other spin-glass
systems, are in Refs.
\cite{Gingras1,Migliorini,Gingras2,Heisenberg,Guven,Ohzeki,Ilker1,Ilker2,Ilker3,Demirtas}.

\section{Global Phase Diagram:\\ Multiple Spin-Glass Phases}

Figs. 1 show a calculated sequence of phase diagram cross sections
for the left-chiral $(c=0)$, on the upper side, and quenched random
left- and right-chiral $(c=0.5)$, on the lower side, systems with in
both cases quenched random ferromagnetic and antiferromagnetic
interactions.  The system exhibits a disordered phase (D), a
ferromagnetic phase (F), a conventionally ordered (in contrast to
the algebraically ordered for $q=5$) antiferromagnetic phase (A), a
quadrupolar phase (Q), a new "one-step" phase (R), a multitude of
different chiral phases, and four different spin-glass phases
(S$_{Ch}$, S$_{FA}$, S$_Q$, S$_R$) including
spin-glass-to-spin-glass phase transitions. The ferromagnetic and
different chiral phases accumulate as conventional and
temperature-inverted (abutting to the reentrant
\cite{Cladis,Hardouin,Garland,Netz,Kumari} disordered phase) devil's
staircases \cite{Bak,Fukuda} at their boundary with the disordered
(D) phase.  This accumulation and its multiplicity of intervening
phases occur at all scales of phase diagram space (i.e., at all
magnifications of the phase diagram figure, as for example seen up
to 100-fold calculated magnification in Fig. 4 of \cite{Caglar2}),
which is the definition of a devil's staircase.

Unlike the odd $q$ case of $q=5$, which incorporates built-in
entropy \cite{Caglar2} even without any quenched randomness, no
algebraically ordered phase \cite{BerkerKadanoff1,BerkerKadanoff2}
occurs in this even $q$ case of $q=4$.  The devil's staircases of
the chiral phases is again seen.  Most interestingly, quadrupolar
and "one-step" phases, different spin-glass phases for the first
time in the same phase diagram, and spin-glass-to-spin-glass direct
phase transitions are seen.  The phases and phase boundaries
involving spin glassiness are tracked through the calculated
Lyapunov exponents of their chaos.

In all ordered phases, the renormalization-group trajectories flow
to strong (infinite) coupling. In the ferromagnetic phase, under
renormalization-group transformations, the interaction $V_{ij}(0)$
becomes asymptotically dominant. In the antiferromagnetic phase,
under renormalization-group transformations, the interaction
$V_{ij}(\pi)$ becomes asymptotically dominant.  In the quadrupolar
phase Q, the interactions  $V_{ij}(0)$ and $V_{ij}(\pi)$ become
asymptotically dominant and equal.  Thus, there are two such
quadrupolar phases, namely along the spin directions $\pm x$ and
$\pm y$, with the additional (factorized) trivial degeneracy of
$\pm$ spin direction at each site.  In the new "one-step phase" R,
the interactions $V_{ij}(+\pi/2)$ and $V_{ij}(-\pi/2)$ become
asymptotically dominant and equal.  Thus, in such a phase, the
average local spins can span all spin directions, taking $\pm \pi/2$
steps from one spin to the next in the renormalized systems. The
identification of the distinct chiral phases, each with distinct
chiral pitches, has been explained in Ref. \cite{Caglar2}.

The renormalization-group trajectories starting in the spin-glass
phases, unlike those in the ferromagnetic, antiferromagneric,
quadrupolar, "one-step", and chiral phases, do not have the
asymptotic behavior where at any scale one potential $V(\theta)$ is
dominant. These trajectories of the spin-glass phases asymptotically
go to a strong-coupling fixed probability distribution
$P(\textbf{V}_{ij})$ which assigns non-zero probabilities to a
distribution of $\textbf{V}_{ij}$ values, with no single
$V_{ij}(\theta)$ being dominant. These distributions are shown in
Figs. 2 and 3.  Different asymptotic fixed probability distributions
indicate different spin-glass phases.

Since, at each locality, the largest interaction in
$\{V_{ij}(0),V_{ij}(\pi/2),V_{ij}(\pi),V_{ij}(3\pi/2)\}$ is set to
zero and the three other interactions are thus made negative, by
subtracting the same constant from all four interactions without
affecting the physics, the quenched probability distribution
$P(\textbf{V}_{ij})$, a function of four variables, is actually
composed of four functions $P_\sigma(\textbf{V}_{ij})$ of three
variables, each such function corresponding to one of the
interactions being zero and the other three, arguments of the
function, being negative. Figs. 2 and 3 show the latter functions.

In Fig. 2 for the spin-glass phase S$_{Ch}$, the part of the fixed
distribution, $P_1(\textbf{V}_{ij})$, for the interactions
$\textbf{V}_{ij}$ in which $V_{ij}(\pi/2)$ is maximum and therefore
0 (and the other three interactions are negative) is shown. The
projections of $P_1(\textbf{V}_{ij})$ onto two of its three
arguments are shown in each panel of Fig. 2. The other three
$P_\sigma(\textbf{V}_{ij})$ have the same fixed distribution. Thus,
chirality is broken locally, but not globally, just as, in the
long-time studied ferromagnetic-antiferromagnetic spin glasses,
spin-direction symmetry breaking is local but not global (i.e., the
local magnetization is non-zero, the global magnetization is zero).
The asymptotic fixed distribution of the phase S$_{Ch}$, given in
Fig. 2, assigns non-zero probabilities to a continuum of values for
all four interactions
$\{V_{ij}(0),V_{ij}(\pi/2),V_{ij}(\pi),V_{ij}(3\pi/2)\}$.  The phase
S$_{Ch}$ is therefore a chiral spin-glass phase.  The similar chiral
spin-glass phase has been seen previously, but as the sole
spin-glass phase, for the odd $q=5$.\cite{Caglar2}.  The chiral
spin-glass phase occurs even when there is no competing
ferromagnetic-antiferromagnetic interactions.\cite{Caglar2,Caglar}

As seen in Fig. 3, in the asymptotic fixed distribution of the
spin-glass phase S$_{FA}$, non-zero probabilities are assigned to a
continuum of values of $\{V_{ij}(0),V_{ij}(\pi)\}$.  Fig. 3 shows
the fixed distribution values $P_0(V_{ij}(\pi))$ for $V_{ij}(0)$
maximum and therefore set to zero. Completing the asymptotic fixed
distribution of S$_{FA}$ is an identical function $P_2(V_{ij}(0))$
for $V_{ij}(\pi)$ maximum and therefore set to zero. At this fixed
distribution, the values of $V_{ij}(\pi/2)$ and $V_{ij}(3\pi/2)$
diverge to negative infinity, so that these angles do not occur.
Thus, S$_{FA}$ is the long-studied \cite{NishimoriBook} spin-glass
phase of competing ferromagnetic and antiferromagnetic interactions.

Fig. 3 also shows the asymptotic fixed distribution of the
spin-glass phase S$_{R}$, with the functions $P_1(V_{ij}(3\pi/2))$
for $V_{ij}(\pi/2)$ maximum (and therefore set to zero) and
$P_3(V_{ij}(\pi/2))$ for $V_{ij}(3\pi/2)$ maximum (and therefore set
to zero). Again, the other two angles do not occur at this
asymptotic fixed distribution. Furthermore, Fig. 3 also shows the
asymptotic fixed distribution of the spin-glass phase S$_{Q}$, with
the functions $P_0(V_{ij}(\pi/2))$ and $P_1(V_{ij}(0))$, with
$V_{ij}(0) = V_{ij}(\pi)$ and $V_{ij}(\pi/2) = V_{ij}(3\pi/2)$.
Thus, this fixed distribution does not locally distinguish between
$\pm$ spin directions and is thus a quadrupolar spin-glass phase.

In fact, the $v_{ij}(0) = v_{ij}(p)$ curve obtained from the left
panel of Fig. 2 also matches the curve here. The three fixed
distributions given in Fig. 3 exhibit the same numerical curve, but
refer to widely different interactions. Thus, they underpin
different spin-glass phases.

\section{Phase Transitions between Chaos}

Another distinctive mechanism, that of chaos under scale change
\cite{McKayChaos,McKayChaos2,BerkerMcKay} or, equivalently
\cite{Ilker1}, chaos under spatial translation, occurs within the
spin-glass phase and differently at the spin-glass phase boundary
\cite{Ilker1}, in systems with competing ferromagnetic and
antiferromagnetic interactions \cite{McKayChaos,
McKayChaos2,BerkerMcKay,Bray,Hartford,
Nifle1,Nifle2,Banavar,Frzakala1,Frzakala2,Sasaki,Lukic,Ledoussal,Rizzo,
Katzgraber,Yoshino,Pixley,Aspelmeier1,Aspelmeier2,Mora,Aral,Chen,Jorg,Lima,Katzgraber2,MMayor,ZZhu,Katzgraber3,Fernandez,Ilker1,Ilker2,MMayor2}
and, more recently, with competing left- and right-chiral
interactions \cite{Caglar, Caglar2}.  Originally found in
hierarchical systems \cite{McKayChaos,McKayChaos2,BerkerMcKay},
scaling or equivalently translation spin-glass chaos is now well
accepted for real $d=3$ lattices and experimental systems
\cite{McKayChaos, McKayChaos2,BerkerMcKay,Bray,Hartford,
Nifle1,Nifle2,Banavar,Frzakala1,Frzakala2,Sasaki,Lukic,Ledoussal,Rizzo,
Katzgraber,Yoshino,Pixley,Aspelmeier1,Aspelmeier2,Mora,Aral,Chen,Jorg,Lima,Katzgraber2,MMayor,ZZhu,Katzgraber3,Fernandez,Ilker1,Ilker2,MMayor2}.

Fig. 4 gives the asymptotic chaotic renormalization-group
trajectories of the four different spin-glass phases and of the
phase boundaries between the spin-glass phases with other spin-glass
phases, with the non-spin-glass ordered phases and the disordered
phase.

Chaos is measured by the Lyapunov exponent, whose calculation for
the multiinteraction
$V_{ij}(0),V_{ij}(\pi/),V_{ij}(\pi),V_{ij}(3\pi/2)$ case is given in
Ref. \cite{Caglar2}. Spin-glass chaos occurs for $\lambda > 0$
\cite{Aral} and the more positive $\lambda$, the stronger is chaos.
Inside all four spin-glass phases, the average interaction diverges
as $<|V|> \sim b^{y_R n}$, where $n$ is the number of
renormalization-group iterations and $y_R = 0.25$ is the runaway
exponent.  In the non-spin-glass ordered phases, the runaway
exponent value is $y_R = d-1 = 3$ \cite{Berker}.

At the S$_{Ch}-$S$_R$, S$_{Ch}-$S$_Q$, S$_{FA}-F$ and its symmetric
S$_{FA}-$A phase boundaries, $y_R = 0.25$ also.  At the
S$_{Ch}-$S$_{FA}$ phase boundary, $y_R = 0.11$ for $V_{ij}(0),
V_{ij}(\pi)$ and $y_R = 0.25$ for $V_{ij}(\pi/2), V_{ij}(3\pi/2)$.
At the phase boundaries of the spin-glass phases with some
non-spin-glass ordered and disordered phases, the average
interaction remains non-divergent, fixed at $<V> = -0.34$ for
S$_{FA}-$Q, S$_R-$R, S$_Q-$D and $<V> = -1.07$ for S$_{Ch}-$D. As
indicated by the Lyapunov exponents, chaos is stronger inside the
spin-glass phase than at its phase boundaries with non-spin-glass
phases.

As expected from the asymptotic fixed distribution analysis given
above, the three spin-glass phases S$_{FA}$, S$_Q$, S$_R$ and the
phase transitions between these phases have the same Lyapunov
exponent $\lambda=1.92$ and therefore the same degree of chaos.  The
chiral spin-glass S$_{Ch}$ has more chaos ($\lambda=1.98$) from the
other three spin-glass phases.  The phase transition between the
chiral spin-glass phase S$_{Ch}$ and the other three spin-glass
phases is a phase transition between different types of chaos.  This
phase transition itself of course exhibits chaos, as do all
spin-glass phase boundaries.

\section{Conclusion}

The left-right chiral and ferromagnetic-antiferromagnetic double
spin-glass clock model, with the crucially even number of states
$q=4$ and in three dimensions $d=3$, has been solved by
renormalization-group theory that is approximate for the cubic
lattice and exact for the corresponding hierarchical lattice.  We
find in the same phase diagram, for the first time to our knowledge,
four different spin-glass phases, including conventional, chiral,
and quadrupolar spin-glass phases, and phase transitions between
spin-glass phases. The chaoses, in the different spin-glass phases
and in the phase transitions of the spin-glass phases with the other
spin-glass phases, the non-spin-glass ordered phases, and the
disordered phase, are determined and quantified by Lyapunov
exponents.  It is seen that the chiral spin-glass phase is the most
chaotic spin-glass phase. The calculated phase diagram is also
otherwise very rich, including regular and temperature-inverted
devil's staircases and reentrances.

The recently found chiral spin-glass phase could possibly be seen in
quenched random dimolecular crystals.  In fact, if magnetic moments
could be included into the component chiral molecules, the double
spin-glass system, with the multiplicity of spin-glass phases seen
here, could be achieved.

\begin{acknowledgments}
Support by the Academy of Sciences of Turkey (T\"UBA) is gratefully
acknowledged.
\end{acknowledgments}

\end{document}